\documentclass [pra, twocolumn, aps, showpacs] {revtex4}
\usepackage{graphicx}
\usepackage{amsmath}
\usepackage{amssymb}

\DeclareMathOperator{\tr}{Tr}

\begin{document}

\title{Two-body recombination in a quantum mechanical lattice gas: Entropy generation and probing of short-range magnetic correlations}

\author{Stefan K. Baur}
\affiliation{Laboratory of Atomic and Solid State Physics, Cornell University, Ithaca, NY 14853, USA}
\email{skb37@cornell.edu}
\author{Erich J. Mueller}
\affiliation{Laboratory of Atomic and Solid State Physics, Cornell University, Ithaca, NY 14853, USA}

\date{\today}

\begin{abstract}
We study entropy generation in a one-dimensional (1D) model of bosons in an optical lattice experiencing two-particle losses.  Such heating is a major impediment to observing exotic low temperature states, and ``simulating" condensed matter systems.  Developing intuition through numerical simulations, we present a simple empirical model for the entropy produced in this 1D setting.  We also explore the time evolution of one and two particle correlation functions, showing that  they are robust against two-particle loss. Because of this robustness, induced two-body losses can be used as a probe of short range magnetic correlations.
\end{abstract}



\maketitle

\begin{figure*}[tbh]
	\centering
		\includegraphics[width=\textwidth]{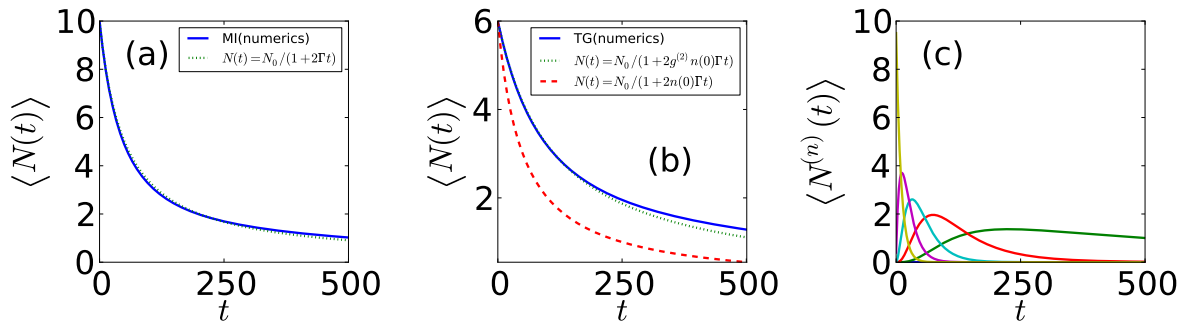}
	\caption{(Color online) Left: Average particle number $\langle N\rangle=\sum_n n \tr \rho^{(n)}$ as a function of time for an initial Mott insulator state on a $L=10$ lattice. Solid line: numerical simulation; Dotted line:  two-body decay law for an uncorrelated state  $N(t)=N(0)/\left[1+2 \Gamma t\right]$. Middle: Same, but for a Tonks-Girardeau gas initial state (ground state of a hard core lattice gas with $L=10,N=6$). Solid line: simulation.  Green dashed curve: two-body decay law for an uncorrelated state. $N(t)=N(0)/(1+2 \Gamma t)$, Dotted line: two-body decay low assuming time independent correlations  $N(t)=N(0)/(1+2 g^{(2)}(0) n(0) \Gamma t)$. Right: Average particle numbers in the different sectors $\langle N^{(n)}(t) \rangle= n \tr \rho^{(n)}(t)$ for the Mott insulator initial state. The sum of all curves at a certain time gives the blue in the leftmost figure.  All times measured in units of the inverse hopping $J^{-1}$.}
	\label{fig:decay}
\end{figure*}

\section{introduction}
As cold gas experimentalists turn their attention toward more strongly correlated states (such as Mott insulators and fractional quantum Hall states) they find that equilibration times become large. This is problematic: inelastic processes limit the time over which one can conduct an experiment, and the experimentalists find themselves in a race. 
Can the system equilibrate before the entropy generated by the inelastic processes destroy the state of interest?  A similar issue arises when  the experimental protocol involves some sort of ``adiabatic" change of parameters (such as ramping up the intensity of an optical lattice): can the adiabatic ramp be completed before inelastic processes take over?  Here we explore a simple  one--dimensional (1D) model where one can quantitatively study the entropy generated by two-body losses.  Within this model we find that the entropy produced by each recombination event is of order the logarithm of the number of atoms, highlighting the difficulty faced by future experiments.  We hope that quantitative studies of such inelastic processes can help overcome them -- though we see no simple solution at this point and time.  On a more positive note, we find that within our model the inter-atomic correlations are largely time independent, even in the presence of drastic atom loss.  Thus even with atom loss one can confidently measure the inter-atomic correlations of an initial state of interest.  While one hopes that this result is generic, it is possible that the robustness of the correlations is an artifact of 1D, where the dynamics are non-ergodic. To test this we have looked at small 2D clusters, finding nearly identical results.  Extrapolating to more complex systems, we give an explicit example of how one could use losses to measure nearest neighbor antiferromagnetic correlations in a two component Fermi gas.

The problem of heating due to atom losses goes back to the first attempts to experimentally create degenerate Fermi gases~\cite{DeMarco1999,Timmermans2001}.  Then it was pointed out that while atom losses are relatively benign in a Bose condensate, they have drastic consequences for a degenerate Fermi gas.  Randomly removing particles from a Fermi sea generates large amounts of entropy.  

Here we consider a version of a model introduced by Verstraete {\it et al.} \cite{Verstraete2004}. This and related models were explored in a number of theoretical and experimental works \cite{Verstraete2004,Kraus2008,Diehl2008,Garcia-Ripoll2009,Daley2009,Durr2009,Kantian2009,Syassen2008,Roncaglia2010}.  The original model consists of a gas of bosons moving in a 1D lattice.  Whenever two bosons are on the same site they recombine with rate $\Gamma_0$.  The composite object which they form is then lost.  This model could describe a gas of molecules (where a recombination mechanism always exists) or a gas of atoms (where light assisted collisions provide a recombination mechanism).  Most experiments are engineered to minimize these two-body losses. They can itentionally be made stronger \cite{Junker2008} and also occur in near-resonant optical traps, such as the blue-detuned lattices used by Schneider {\it et al.} ~\cite{Schneider2008}.


When $\Gamma_0$ is sufficiently high, or in the presence of strong on-site interactions, one can integrate out the doubly occupied sites, producing a model with hard-core interactions and a nearest neighbor loss term.  Mathematically the time evolution of this dissipative system is then given by a Master equation
\begin{eqnarray}\label{master}
\frac{d \rho}{dt}&=&-\frac{i}{\hbar} [H,\rho]\\&+&\Gamma \sum_{\langle i,j \rangle}\left[ a_i a_j \rho a_j^{\dagger} a_i^{\dagger}-\frac{1}{2} \left( n_i n_j \rho+\rho n_i n_j \right) \right], \nonumber
\end{eqnarray}
where the conservative part of the dynamics are described by the Hamiltonian
\begin{eqnarray}
H=-J \sum_{\langle i,j \rangle} a_i^{\dagger} a_j+a_j^{\dagger} a_i.
\end{eqnarray}
In these equations, $a_i$ is the operator which annihilates an atom at site $i$, $J$ is the tunneling matrix element, $\rho$ is the density matrix, $\hbar$ is the reduced Planck's constant, $\langle i,j\rangle$ denotes nearest neighbor sites, $1/\Gamma$ is the time it takes for atoms at nearest neighbor sites to recombine.  These equations implicitly assume that one works in a Hilbert space where each site is occupied by only zero or one particle.  The remarkable result found in previous work is that $\Gamma$ scales inversely  $\Gamma_0$.  Thus when $\Gamma_0\to\infty$, the dynamics become conservative, coinciding with those of a hard-core gas of particles, a lattice Tonks-Girardeau gas.

We are interested in how entropy and correlations evolve with time in this model when $\Gamma\ll J$, but $\Gamma\neq0$.  For example, suppose one begins at time $t=0$ in an $N$-particle Tonks state.  As time evolves atoms are lost, until the system has $N^\prime<N$ particles.  Does the system adiabatically evolve into a $N^\prime$ particle Tonks gas (or an ensemble of such gases with a different particle numbers)?  We find that this is not the case.  As we describe below, we find that the $N^\prime$ particle system is better described by the initial $N$-particle Tonks state with random atoms removed.  This is a high entropy state. Despite its highly non-equilibrium character, it inherits the two-particle correlations of the initial $N$-particle Tonks gas.
These correlations, which are very different from what one expects for the $N'$-particle Tonks gas, are directly measurable, and greatly impact the behavior of the system.

\section{Numerical Approach}
To solve the Master equation in Eq.~(\ref{master}), we consider a small chain of length $L$.   We numerate all $2^L$ possible many-body states in which no more than one particle sits on each site.   We explicitly write the density matrix in this basis as a $2^L\times 2^L$ matrix, and express Eq.~(\ref{master}) as a coupled system of equations for the $2^{2L}$ matrix elements.   Note the time evolution does not create coherences between states of different particle number.  This lack of coherence represents the fact that recombination events at different places or time are in principle distinguishable.  Consequently the density matrix is block diagonal, and for even number of particles $N$ can be written
\begin{eqnarray}
\rho=\rho^{(0)}\oplus \rho^{(2)} \oplus \ldots \oplus \rho^{(N)}
\end{eqnarray}
where $\rho^{(n)}$ is the $L \choose n$-dimensional density matrix for the sector with $n$ particles.  We use a split-step method for our time evolution --- alternating the exact Hamiltonian dynamics with the exact dissipative dynamics.  We vary our time step to verify that our results are independent of the time step.  We find that it is impractical to take $L>12$, as the Hilbert space becomes too large. We typically quote results using $L=10$ or $L=12$.  

Other approaches, such as the density matrix renormalization group can be applied to this problem \cite{Garcia-Ripoll2009}, allowing one to consider larger systems, but making some observables more difficult to calculate

\begin{figure*}[tbh]
	\centering
		\includegraphics[width=0.7\textwidth]{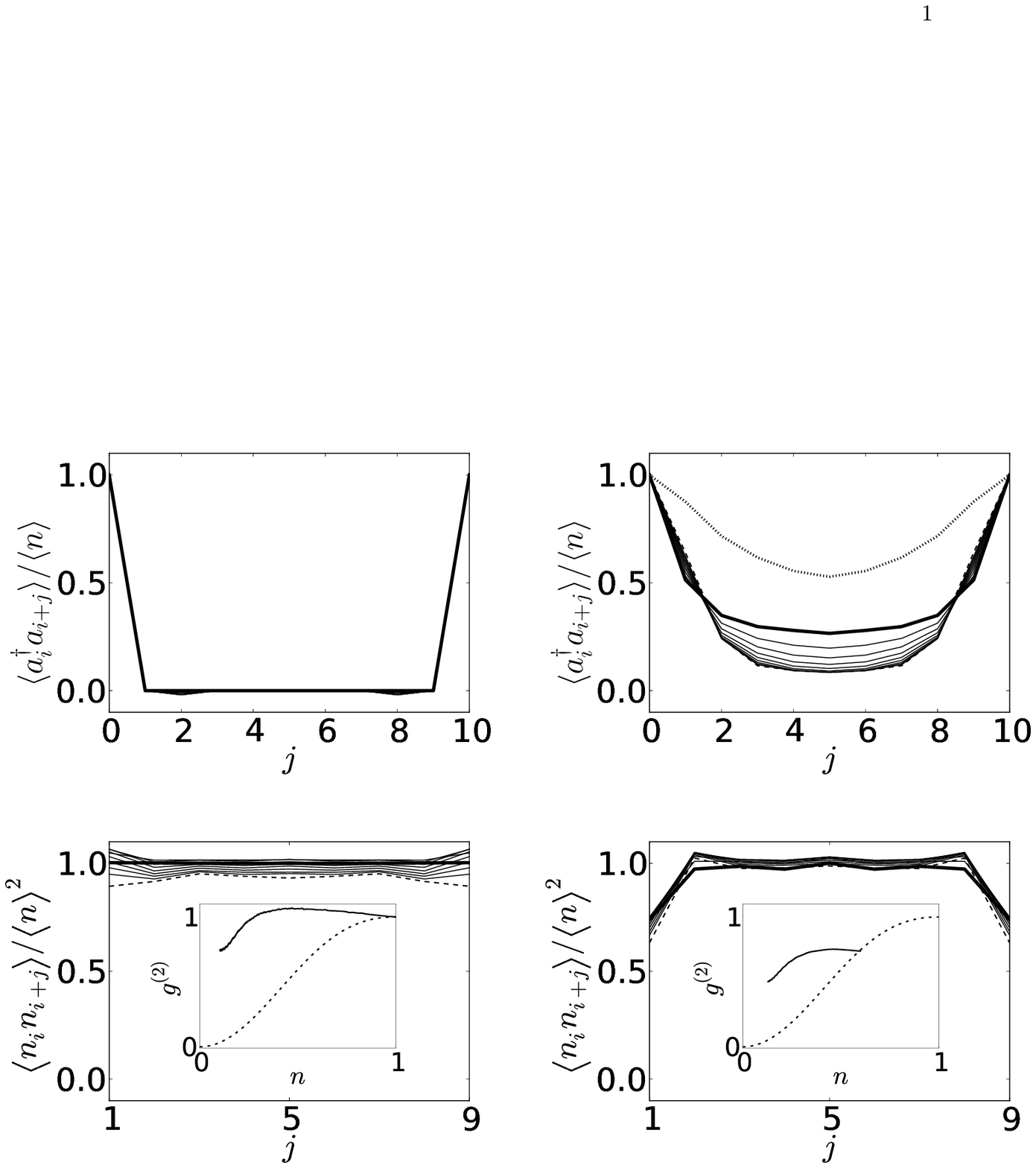}
	\caption{Time evolution of correlation functions starting from (left) the 10 particle Mott Insulator ($L=10,N=10$) or (right) the 6 particle Tonks-Girardeau state ($L=10,N=6$).  Thick line: $t=0$; Dashed line: $t=200 J^{-1}$; Thin lines: intermediate times separated by $20 J^{-1}$; Dotted line: The single particle density matrix $\langle a_i^{\dagger} a_{i+j}\rangle$ one would expect if each of the n-particle sectors were in their ground state at $t=200 J^{-1}$. The insets of the lower-left and lower-right figures show $g^{(2)}$ as a function of density $n=N/L$ together with the analytic formula for an infinite hardcore boson system in the ground state at the same density $g^{(2)}_{\rm eq}(n)=1-[\sin(\pi n)/(n \pi)]^2$. 	}
	\label{fig:cor}
\end{figure*}

\section{Time evolution of the density}
As described by Garc\'ia-Ripoll et al. \cite{Garcia-Ripoll2009}, the time evolution of the density in this model can be understood by a rather simple argument.  One begins by noting that the rate of change of the number of atoms at a site only depends on the correlations between particles on nearby sites:
\begin{eqnarray}
\frac{d \langle n_i \rangle}{dt}=- \Gamma \left[\langle n_i n_{i+1}\rangle+\langle n_i n_{i-1}\rangle\right]
\end{eqnarray}
Translational invariance implies that the two terms in brackets are equal to one-another, and one can write  
\begin{eqnarray}
\frac{dn}{dt}=-2 \Gamma g^{(2)} n^2(t)
\end{eqnarray}
where $g^{(2)}=\langle n_{i} n_{i+1}\rangle/ \langle n_{i}\rangle^2$ measures pair correlations and is related to the probability for finding two particles on neighboring sites. (Note: this differs from the definition of $g^{(2)}$ in \cite{Syassen2008}.)  For uncorrelated sites, such as one finds in the Mott insulator,  one has $g^{(2)}=1$. 
As we will directly illustrate in Sec. ~\ref{seccor}, when $n<1$, the exact
ground state of $H$ has $g^{(2)}<1$, and these equilibrium correlations are strongly number dependent.  Despite this behavior of the equilibrium correlations,  the time evolution in Eq.~(\ref{master})  leaves the initial correlations nearly unchanged: $g^{(2)}(t)\approx g^{(2)}(t=0)$.  The degree to which this holds will be quantified below.  If one treats the correlations as static  one finds
\begin{equation}\label{simpdecay}
n(t)=\frac{n(0)}{1+2 t \Gamma  n(0) g^{(2)} (0)}.
\end{equation}  
Variations on this equation have appeared in the literature \cite{Garcia-Ripoll2009, Syassen2008}, which can be interpreted as different models for $g^{(2)}(t)$.

Figure~\ref{fig:decay} (a,b) compares the time evolution found from the Master equation with that from Eq.~(\ref{simpdecay}).  Panel (a) shows the evolution beginning from a Mott state (with $g^{(2)}(t=0)=1$), while panel (b) shows the decay of a correlated state.  
Panel (c) shows the weight in each of the different particle number channels.

Similar results for the decay of the Mott insulator were found in \cite{Syassen2008}.  In particular, Garc\'ia-Ripoll {\it et al.} \cite{Garcia-Ripoll2009} used a Density Matrix Renormalization Group simulation to show that the observations in Fig.~1(a) are independent of system size.  In the remainder of this paper we extend these results, directly analyzing the time dependence of correlations and entropy.

\section{Time evolution of two-site obeservables}\label{seccor}
We quantify the time evolution of the correlations by studying two objects: the single particle density matrix $\langle a_i^\dagger a_{i+j}\rangle/\langle n \rangle$, and the density-density correlations function $\langle n_i n_{i+j}\rangle/\langle n\rangle^2$.  The equilibrium value of the latter correlation function is efficiently calculated by performing a Jordan-Wigner transformation and mapping the hard-core Bose gas onto a gas of non-interacting Fermions. 
In particular, in the thermodynamic limit the equilibrium nearest neighbor density-density correlator at density $n$ is
\begin{equation}\label{eqg2}
g_{\rm eq}^{(2)}(n)=1-\left(\frac{\sin \pi n}{\pi n}\right)^2.
\end{equation}
As already emphasized, we see large deviations from this equilibrium prediction.

Figure~\ref{fig:cor} shows the time dependence of the correlation functions for the two initial conditions previously explored.  The two figures on the left show the behavior of the Mott state.  Not only are the density correlations largely time independent, but so is the single-particle density matrix.    One immediate implication is that 
 the atom momentum distribution (and hence a time-of-flight image) is unchanged by the loss.  The inset of the lower left figure compares the $g^{(2)}$ extracted from our simulations to Eq.~(\ref{eqg2}).
 
 As shown by the two figures on the right, the dynamics from the Tonks state also leads to nearly time independent correlation functions.
 As time evolves there is a very slight drop in the nearest neighbor density correlations, and the single particle density matrix begins to fall off more rapidly with distance.
This redistribution of the off-diagonal weight of the single particle density matrix corresponds to a shift of particles to larger momentum. Interestingly, this is the opposite of what one would expect if one instead modeled the dynamics as just an adiabatic change in the number of particles.  The inset to the top-right figure shows the equilibrium single particle density matrix (corresponding to the  number of particles are at time $t=200 J^{-1}$).  The slower 
slower spatial variation of the equilibrium $\langle a_i^\dagger a_{i+j}\rangle/\langle n \rangle$, corresponds to a lower occupation of large $k$ states.  This is intuitively sensible, since when $n=1$ one should have only the $k=0$ state occupied.

In addition to being conceptually important, these correlation functions are directly observable.  For example
 a time of flight measurement of the momentum distribution (as was for example done in ~\cite{Paredes2004})  yields the Fourier transform of the single particle density matrix.
The full density-density correlation function can be studied within experiments with single-site resolution such as ~\cite{Bakr2009}. Alternatively noise correlation measurements \cite{Imambekov2009} or elastic light scattering \cite{Corcovilos2010} also probe this static structure factor.

\section{Entropy}
We now proceed to calculate the entropy 
\begin{eqnarray}
S(t)=-\tr \rho(t) \ln \rho(t).
\end{eqnarray}
For ultra-cold atom experiments, where one is dealing with a small isolated system, the entropy is a more relevant than the temperature.  This is especially true here, where the dynamics take one out of thermal equilibrium.

\begin{figure*}[th]
	\centering
		\includegraphics[width=\textwidth]{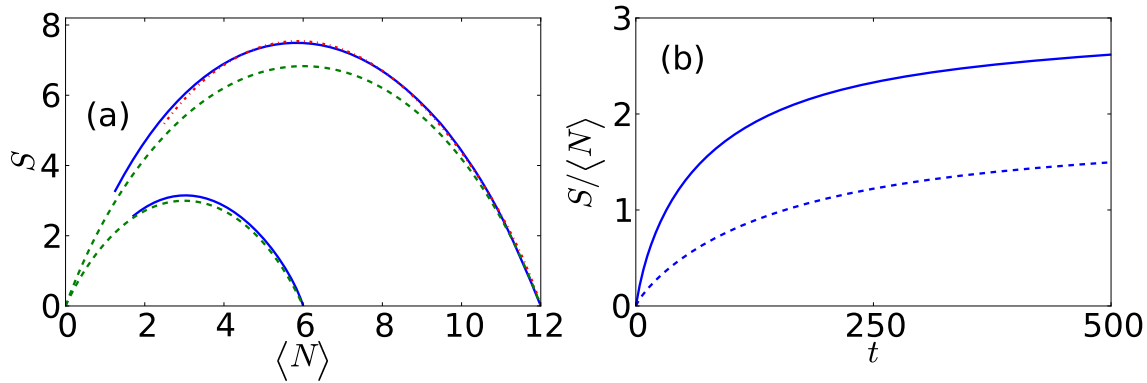}
	\caption{(Wide, Color Online) Left: Entropy $S$ as a function of average particle number $\langle N \rangle$ during time evolution, starting from the (top, solid line) $L=12, N=12$ Mott insulator, (top, dashed-dotted line) $4 \times 3$, $N=12$ (2D) Mott insulator, and (bottom, solid line) $L=12, N=6$ Tonks-Girardeau initial states. Dashed line:  analytic formula $S\sim \ln {N(0) \choose N}$.
	 Right: Entropy per particle as a function of time starting from the (solid line) $L=12$, $N=12$ Mott Insulator and (dashed line) $L=12$, $N=6$ Tonks-Girardeau state with $\Gamma=0.01 J$.
	 }
	\label{fig:entropy}
\end{figure*}

It is convenient, as in Fig.~\ref{fig:entropy}, to parametrically plot $S(t)$ as a function of $N(t)$.  For $\Gamma\ll J$ the resulting curve is then independent of $\Gamma$.  We find that the entropy is well approximated by
 the simple law
\begin{eqnarray}\label{apent}
S(N) \sim \ln{{N_0}\choose{N}},
\end{eqnarray}
where $N_0$ is the number of particles at $t=0$.  

There is a particularly simple interpretation of this result when one starts in the Mott insulating state.  The entropy in Eq.~(\ref{apent}) is what one would find if one randomly punched holes in the Mott insulator.  One would naively expect Eq.~(\ref{apent}) to be an upper bound to the entropy, yet the numerical simulations find an entropy which is strictly above this curve.  The extra entropy principally comes from the fact there is an indefinite number of particles.
When the Mott insulator is depleted to half-filling the entropy is maximal and one has an entropy per particle of (for $L,N \rightarrow \infty$)
\begin{eqnarray}
S/N \sim 2 \ln(2)\approx 1.4.
\end{eqnarray}

A similar interpretation can be produced for the entropy of a depleted Tonks-Girardeau gas.  There it is convenient to map the system onto a gas of non-interacting fermions 
via a Jordan-Wigner transform.  The fermions initially occupy $N_0$ different momentum states.  If one randomly removes fermions from momentum states one arrives at the entropy in Eq.~(\ref{apent}). This is similar to the arguments of Timmermans, where a trapped noninteracting Fermi gas in three dimensions was considered~\cite{Timmermans2001}. Again, the simulation finds an entropy slightly larger than this estimate. 

To understand the applicability of our results to higher dimensional systems, we have performed simulations of a two dimensional (2D) hardcore Bose gas on a rectangular square lattice geometry with a dimension of $4 \times 3$ sites and periodic boundary conditions. The result for entropy vs. particle number is also shown in Fig. \ref{fig:entropy} (a) and is almost identical to what we found in a 1D system with the same number of lattice sites. 

\section{Induced losses as a probe of local spin correlations}
Our observation that initial correlations are preserved during time evolution makes two-body losses an extremely powerful probe of cold atoms.
Losses have long been recognized as a probe of correlations, but have always been viewed as quite invasive~\cite{Kinoshita2005,Syassen2008,Jo2009,Popp2004,Sarajlic2008,Roncaglia2010}.
We find that
  two-body losses are a reliable way to learn about the {\em initial} correlations, even when a large fraction of the atoms are depleted.

As a particular example, 
 we now describe how photoassociation induced losses~\cite{Rom2004,Kinoshita2005,Junker2008} may be used to measure short-range magnetic correlations in two-species lattice bosons or fermions in the $n=1$ Mott insulating phase~\cite{Weld2009}. This approach complements methods that detect long range order, such as noise correlations and light scattering~\cite{Altman2004,Corcovilos2010}. Section ~\ref{fermions} will discuss the fermionic case, while Sec.~\ref{boson} will deal with bosons.
 
 We wish to emphasize that this weak photassociation approach is very different from a {\em sudden} probe such as sweeping the system through a Feshbach resonance or photoassociating the system using a STIRAP protocol \cite{Schneider2008,Mackie2000}.  In those latter approaches the signal size is limited by the instantaneous number of doubly occupied sites, which scales as $J/U$.  For the weak probes used here, however, one can remove a large fraction of the atoms.

The nearest neighbor spin correlations are a smooth function of temperature~\cite{Chiesa2010}, and represent an important precursor of the magnetic order which sets in on temperatures $T\sim J_s$, where $J_s$ is the coupling constant in the effective spin model. Alternative approaches to measure these correlations use lattice modulation spectroscopy ~\cite{Kollath2006} or manipulation of double well potentials combined with band mapping~\cite{Bloch2010}.

\subsection{Two species fermions}\label{fermions}
We consider a two-species Fermi gas in a deep optical lattice that can be described by a Hubbard-model
\begin{eqnarray}
H=-J \sum_{\langle i,j \rangle,\sigma=\uparrow,\downarrow}\left( c_{i\sigma}^{\dagger} c_{j\sigma}+c_{j\sigma}^{\dagger} c_{i\sigma}\right)+U \sum_i n_{i,\uparrow} n_{i,\downarrow}
\end{eqnarray}
We envision introducing a photoassociation laser which drives two atoms on the same site into a molecular state, which is lost from the system.  The bare loss rate $\Gamma_0$, depends on the intensity of the photoassociation laser, as well as details of the atomic/molecular states.
Integrating out the doubly-occupied sites produces a master equation similar to Eq.~(\ref{master}), which can  formally be represented as a a complex Hamiltonian
\begin{eqnarray}
H_F=\left( \tilde{J}_{F}+i 4 \Gamma \right) \sum_{\langle i,j\rangle} \left( \mathbf{S}_i \cdot \mathbf{S}_j- \frac{1}{4} n_i n_j \right)
\end{eqnarray}
with $\tilde{J}_{F}=4 J^2 U_0/\left[U_0^2+(\Gamma_0/2)^2)\right]$ and $\Gamma=J^2 (\Gamma_0/2)/\left[U_0^2+(\Gamma_0/2)^2\right]$.
We envision letting the system equilibrate with the photoassociation lasers turned off ($\Gamma_0=0$).  The lasers are then turned on at a low enough intensity that $\tilde{J}_F \approx J_F$.
The subsequent density evolution will then be described by
\begin{eqnarray}
\frac{dn}{dt}=-q \Gamma g_{\uparrow \downarrow}^{(2)} n^2(t),
\end{eqnarray}
where $q$ is the number of nearest neighbors and the correlation function $g^{(2)}_{\uparrow \downarrow}$ is given by
\begin{eqnarray}\label{result}
g^{(2)}_{\uparrow \downarrow}=\langle n_i n_j-4 \mathbf{S}_i \cdot \mathbf{S}_j\rangle/\langle n_i \rangle^2.
\end{eqnarray}
The initial two-body loss coefficent is proportional to $g^{(2)}_{\uparrow \downarrow}(t=0)=1-4 \langle \mathbf{S}_i \cdot \mathbf{S}_j\rangle$ and provides a direct measure of nearest neighbor spin correlations.
\subsection{Two species bosons}\label{boson}
For an $n=1$ Mott insulator of a two-species Bose gas one can similarly measure local spin correlation functions. Integrating out doubly occupied sites, the Hamiltonian of such a system is formally \cite{Kuklov2003}
\begin{eqnarray}
H_{B}=\sum_{\langle i,j\rangle} J_z S_i^{z} S_j^{z}-J_{\perp} \left( S_i^x S_j^x+S_i^y S_j^y\right)\\+\frac{h}{2} \left( S_i^z n_j+n_i S_j^z\right)-V n_i n_j. \nonumber
\end{eqnarray}
In the absence of losses, the coupling constants are related to the hopping rate of the two species ($J_{\uparrow},J_{\downarrow}$), and the on-site interactions between same ($U_{\uparrow\uparrow}$, $U_{\downarrow\downarrow}$) and different ($U_{\uparrow\downarrow}$) species,
\begin{eqnarray}
J_{\perp}&=&\frac{4 J_{\uparrow} J_{\downarrow}}{U_{\uparrow \downarrow}} ;\hspace{5mm} h=4 \left[ \frac{J_{\downarrow}^2}{U_{\downarrow\downarrow}}-\frac{J_{\uparrow}^2}{U_{\uparrow\uparrow}}\right],\\
J_z&=&J_{\uparrow}^2 \left[ \frac{2}{U_{\uparrow \downarrow}}-\frac{4}{U_{\uparrow \uparrow}}\right]+J_{\downarrow}^2 \left[ \frac{2}{U_{\uparrow \downarrow}}-\frac{4}{U_{\downarrow \downarrow}}\right],\\
V&=&J_{\uparrow}^2 \left[ \frac{1}{2 U_{\uparrow\downarrow}}+\frac{1}{U_{\uparrow\uparrow}}\right]+J_{\downarrow}^2 \left[ \frac{1}{2 U_{\uparrow\downarrow}}+\frac{1}{U_{\downarrow \downarrow}}\right].
\end{eqnarray}
 With a two species Bose gas one can selectively address three different photoassociation transitions($\uparrow\uparrow \rightarrow $molecule, $\uparrow\downarrow \rightarrow $molecule and $\downarrow\downarrow \rightarrow $molecule). This versatility may be used to measure a variety of nearest-neighbor spin-correlation functions. For example, driving a photoassociation resonance that converts an $\uparrow$- and a $\downarrow$-boson into a molecule can be formally described by substituting $U_{\uparrow\downarrow}\rightarrow U_{\uparrow\downarrow}-i \Gamma^0_{\uparrow\downarrow}/2$. 
Specializing to the case $J_{\uparrow}=J_{\downarrow}=J$, 
corresponding to a typical 
optical lattice 
setup,
 one has
\begin{eqnarray}
H&=&\sum_{\langle i,j\rangle}  \left[ \tilde{J}_z S_i^z S_j^z-\tilde{J}_{\perp} \left( S_i^x S_j^x+S_i^y S_j^y \right) \right. \\   &+&\left.\frac{h}{2} \left( S_i^z n_j+n_i S_j^z\right)-\tilde{V} n_i n_j \right]\nonumber \\&-&i 4 \Gamma_{\uparrow \downarrow} \sum_{\langle i,j \rangle} \left[\frac{1}{4} n_i n_j-S_i^z S_j^z+S_i^x S_j^x+S_i^y S_j^y\right] \nonumber
\end{eqnarray}
where $\tilde{J}_{\perp}=4 J^2 U_{\uparrow \downarrow}/[U_{\uparrow \downarrow}^2+(\Gamma^0_{\uparrow \downarrow}/2)^2]$, $\tilde{J}_z=4 J^2 U_{\uparrow \downarrow}/[U_{\uparrow\downarrow}^2+(\Gamma^0_{\uparrow \downarrow}/2)^2]-4 J^2 (1/U_{\uparrow \uparrow}+1/U_{\downarrow \downarrow})$, $\tilde{V}=J^2/[U_{\uparrow \downarrow}^2+(\Gamma_{\uparrow\downarrow}/2)^2]+J^2(1/U_{\uparrow \uparrow}+1/U_{\downarrow \downarrow})$ and $\Gamma_{\uparrow \downarrow}= (\Gamma^0_{\uparrow \downarrow}/2) J^2/[U_{\uparrow \downarrow}^2+(\Gamma_{\uparrow \downarrow}/2)^2]$.
In this case two-body losses measure
\begin{eqnarray}\label{result2}
g^{(2)}_{\uparrow \downarrow}=\langle n_i n_j - 4 (S_i^z S_j^z -S_i^x S_j^x - S_i^y S_j^y)\rangle/\langle n_i \rangle^2
\end{eqnarray}
where $i,j$ are nearest neighbors. Alternatively one can photoassociate $\uparrow\uparrow \rightarrow$molecule to measure $g^{(2)}_{\uparrow \uparrow}=\langle n_i n_j+4 S_i^z S_j^z+2(S_i^z n_j+n_i S_j^z)\rangle/\langle n_i \rangle^2 $.  Photoassociating $\downarrow\downarrow \rightarrow$molecule flips the sign of $S_z$, giving $g^{(2)}_{\downarrow \downarrow}=\langle n_i n_j+4 S_i^z S_j^z-2(S_i^z n_j+n_i S_j^z)\rangle/\langle n_i \rangle^2 $. Simultaneously photoassociating $\uparrow\uparrow$ and $\downarrow\downarrow$ at the same rate allows measurement of $(g_{\uparrow \uparrow}^{(2)}+g_{\downarrow \downarrow}^{(2)})/2=\langle n_i n_j+4 S_i^z S_j^z\rangle$.

For a Mott insulating state of two species bosons, the expression in Eq.~(\ref{result2}) simplifies to
$g^{(2)}_{\uparrow \downarrow}(t=0)=1-4\langle S_i^z S_j^z -S_i^x S_j^x - S_i^y S_j^y\rangle$. Linear combinations of $g^{(2)}_{\uparrow \downarrow},g^{(2)}_{\uparrow \uparrow},g^{(2)}_{\downarrow \downarrow}$ probe $\langle \mathbf{S}_i \cdot \mathbf{S}_j\rangle$ , $\langle S_i^z S_j^z \rangle$ and $\langle S_i^x S_j^x+S_i^y S_j^y\rangle$. 

Generically, for both bosons and fermions, the spin correlations at low temperatures tend to increase the loss rate.  The ferromagnetic super-exchange in a Bose system encourages same-species atoms to sit next to each other, which due to Bose enhancement leads to an increased probability of doubly-occupying a site.  The antiferromagnetic super-exchange in a Fermi system enhances the probability of up-spins lying beside down-spins, increasing the chance that they will end up on the same site.



\section{ACKNOWLEDGMENTS}
We would like to thank Mukund Vengalatorre, Randy Hulet, Brian DeMarco, Stefan Natu, Kaden Hazzard and Jildou Baarsma for useful discussions. This work was supported under ARO Award W911NF-07-1-0464 with funds from the DARPA OLE Program.

\end{document}